\documentclass[twocolumn]{aastex631}

\usepackage{booktabs}
\usepackage{CJKutf8}

\shorttitle{LkCa~15 Horseshoe Ring}
\shortauthors{Long et al.}

\graphicspath{{./}{figures/}}
\begin{document}
\begin{CJK*}{UTF8}{gbsn}

\title{ALMA Detection of Dust Trapping around Lagrangian Points in the LkCa~15 Disk}

\correspondingauthor{Feng Long}
%\email{feng.long@cfa.harvard.edu}
\email{fenglong@email.arizona.edu}

\author[0000-0002-7607-719X]{Feng Long(龙凤)}
\affiliation{Center for Astrophysics \textbar\, Harvard \& Smithsonian, 60 Garden St., Cambridge, MA 02138, USA}

\author[0000-0003-2253-2270]{Sean M. Andrews}
\affiliation{Center for Astrophysics \textbar\, Harvard \& Smithsonian, 60 Garden St., Cambridge, MA 02138, USA}

\author[0000-0002-8537-9114]{Shangjia Zhang(张尚嘉)}
\affiliation{Department of Physics and Astronomy, University of Nevada, Las Vegas, 4505 S. Maryland Pkwy, Las Vegas, NV 89154, USA}

\author[0000-0001-8642-1786]{Chunhua Qi}
\affiliation{Center for Astrophysics \textbar\, Harvard \& Smithsonian, 60 Garden St., Cambridge, MA 02138, USA}

\author[0000-0002-7695-7605]{Myriam Benisty}
\affiliation{Univ. Grenoble Alpes, CNRS, IPAG, 38000 Grenoble, France}
\affiliation{Universit\'e C\^ote d'Azur, Observatoire de la C\^ote d'Azur, CNRS, Laboratoire Lagrange, France}

\author[0000-0003-4689-2684]{Stefano Facchini}
\affiliation{Universit\`a degli Studi di Milano, via Giovanni Celoria 16, 20133 Milano, Italy}

\author[0000-0001-8061-2207]{Andrea Isella}
\affiliation{Department of Physics and Astronomy, Rice University, 6100 Main Street, MS-108, Houston, TX 77005, USA}

\author[0000-0003-1526-7587]{David J. Wilner}
\affiliation{Center for Astrophysics \textbar\, Harvard \& Smithsonian, 60 Garden St., Cambridge, MA 02138, USA}

\author[0000-0001-7258-770X]{Jaehan Bae}
\affiliation{Department of Astronomy, University of Florida, Gainesville, FL 32611, USA}

\author[0000-0001-6947-6072]{Jane Huang}
\altaffiliation{NASA Hubble Fellowship Program Sagan Fellow}
\affiliation{Department of Astronomy, University of Michigan, 323 West Hall, 1085 S. University Avenue, Ann Arbor, MI 48109, USA}

\author[0000-0002-8932-1219]{Ryan A. Loomis}
\affiliation{National Radio Astronomy Observatory, 520 Edgemont Rd., Charlottesville, VA 22903, USA}

\author[0000-0001-8798-1347]{Karin I. \"Oberg} 
\affiliation{Center for Astrophysics \textbar\, Harvard \& Smithsonian, 60 Garden St., Cambridge, MA 02138, USA} 

\author[0000-0003-3616-6822]{Zhaohuan Zhu(朱照寰)}
\affiliation{Department of Physics and Astronomy, University of Nevada, Las Vegas, 4505 S. Maryland Pkwy, Las Vegas, NV 89154, USA}

\begin{abstract}
We present deep high-resolution ($\sim$50\,mas, 8\,au) ALMA 0.88 and 1.3\,mm continuum observations of the LkCa~15 disk. 
The emission morphology shows an inner cavity and three dust rings at both wavelengths, but with slightly narrower rings at the longer wavelength. 
Along a faint ring at 42\,au, we identify two excess emission features at $\sim$10$\sigma$ significance at both wavelengths: one as an unresolved clump and the other as an extended arc, separated by roughly 120 degrees in azimuth. The clump is unlikely to be a circumplanetary disk (CPD) as the emission peak shifts between the two wavelengths even after accounting for orbital motion. Instead, the morphology of the 42\,au ring strongly resembles the characteristic horseshoe orbit produced in planet--disk interaction models, where the clump and the arc trace dust accumulation around Lagrangian points $L_{4}$ and $L_{5}$, respectively. The shape of the 42\,au ring, dust trapping in the outer adjacent ring, and the coincidence of the horseshoe ring location with a gap in near-IR scattered light, are all consistent with the scenario of planet sculpting, with the planet likely having a mass between those of Neptune and Saturn. We do not detect point-like emission associated with a CPD around the putative planet location ($0\farcs27$ in projected separation from the central star at a position angle of $\sim$60\degr), with upper limits of 70 and 33\,$\mu$Jy at 0.88 and 1.3\,mm, respectively, corresponding to dust mass upper limits of 0.02--0.03\,$M_{\oplus}$.

\end{abstract}


\section{Introduction} \label{sec:intro}
Disk substructures, in the forms of gaps and rings, spiral arms, and asymmetries, are prevalent in high angular resolution images from both millimeter continuum emission and optical/near-infrared scattered light (see recent reviews by \citealt{Andrews2020, Benisty2022arXiv}). These features are widely taken as the imprints of planet--disk interactions (e.g., \citealt{Zhu2011, Pinilla2012, Dipierro2015}), predicting a wealth of young planets with masses of 0.1--1\,$M_{\rm Jup}$ at orbital distances of 10--100\,au (e.g., \citealt{Bae2018,Zhang2018,Lodato2019}). However, confirming the presence of these planets remains challenging, as the current detection limits from direct imaging surveys are on the order of a few Jupiter masses (e.g., \citealt{Asensio-Torres2021}).
Only in very few cases, protoplanet detections have been reported (e.g., \citealt{Keppler2018, Currie2022NatAs}), which is consistent with the picture that most planetary perturbers are in the low mass regime (see also \citealt{Brittain2020} for alternative explanation).

Given the challenges of directly detecting young planet photospheres, efforts from various indirect avenues are essential to confirm the planet origin of disk substructures. One recent advance is the detection of localized velocity deviations from Keplerian motion in the gas disk, where the perturbation amplitude is linked to the planet mass \citep{Pinte2022arXiv}. Another option aims to probe disk material that feeds the growth of a planet through a circumplanetary disk (CPD), on scales smaller than the Hill radius. Deep high-resolution millimeter observations could reveal such faint and compact CPDs, as have been demonstrated by the recent CPD detection around PDS\,70\,c \citep{Isella2019, Benisty2021} and a new candidate in the AS\,209 disk \citep{Bae2022arXiv}. A third, rarely explored, feature of planet--disk interactions is the accumulation of disk material around Lagrangian points along the planet orbit \citep[e.g.,][]{Lyra2009, Montesinos2020}. 
For example, the one-sided crescent-shaped asymmetry in the HD\,163296 disk has been suggested to mimic dust trapping in the Lagrangian point $L_5$ \citep{Rodenkirch2021}.

We focus a search for evidence of these planet-disk interaction features in the continuum emission here on the LkCa\,15 system. LkCa\,15 is a K5 star with a mass of $1.2\pm0.1\,M_{\odot}$, located in the Taurus star-forming region at a distance of 159\,pc \citep{Gaia2018, Donati2019}. Early millimeter observations reported a dust-depleted cavity in the disk (out to a radius of $\sim$45\,au; \citealt{Pietu2006, Andrews2011}), making it a popular target for young planet searches. 
With high resolution ($68\times47$\,mas) Atacama Large Millimeter/submillimeter Array (ALMA) observations at 1.3\,mm, \citet{Facchini2020} showed that the disk has three dust rings exterior to the cavity and suggested that a planet is forming around the bright middle ring. 
Three giant planet candidates have been claimed through optical and IR imaging techniques at radial distances of $0\farcs10-0\farcs13$ (16--21 au; \citealt{Kraus2012, Sallum2015}), which however have been later interpreted as inner disk emission \citep{Thalmann2016, Currie2019, Blakely2022}.
No CPDs associated to these candidates were detected in 7\,mm continuum \citep{Isella2014}. 

In this Letter, we combine the long-baseline data presented in \citet{Facchini2020} with new observations and report the discovery of two excess emission features along the faintest ring in the LkCa\,15 disk. Those features resemble the scenario of dust trapping in the $L_4$ and $L_5$ Lagrangian points around an unseen planet at a semimajor orbital radius of 42\,au. We present ALMA observations and the details of data reduction in Section~\ref{sec:obs}. The overall dust emission morphology and the new finding of significant excess emission along one dust ring are reported in Section~\ref{sec:results}. We then discuss these results in the context of planet formation in Section~\ref{sec:diss} and summarize in Section~\ref{sec:summ}.

\begin{deluxetable*}{cccccccc}[!t]
%[!t]
%\rotate
\tabletypesize{\scriptsize}
\tablecaption{ALMA Observation Summary\label{tab:obs}}
\tablewidth{0pt}
\tablehead{
\colhead{Program ID} & \colhead{PI} & \colhead{Obs.Freq.} & \colhead{UTC Time} & \colhead{$N_{\rm ant}$} & \colhead{Baseline} &  \colhead{On-source Time} & \colhead{Ref.Ant}  \\
\colhead{} & \colhead{} & \colhead{(GHz)} & \colhead{}   & \colhead{} & \colhead{(meter)} & \colhead{(min)} & \colhead{}  \\
%(mJy beam$^{-1}$ km s$^{-1}$)
%R$_{gas}$/R$_{mm}$
\colhead{(1)} & \colhead{(2)} & \colhead{(3)} & \colhead{(4)} & \colhead{(5)} & \colhead{(6)} & \colhead{(7)} & \colhead{(8)}
}
\startdata
\multicolumn{8}{c}{Band 6 Observations} \\ 
\hline
2015.1.00118.S & L.~Looney	 & 223.06-242.93 & 2017 July 09-12:08 &	 40	&  16-2647	 &	12  &		DA62  \\	
2015.1.00118.S & L.~Looney	 & 223.06-242.93 & 2017 July 09-13:31 &	 40	&  16-2647	 &	12  &		DA62 	  \\
2015.1.00118.S & L.~Looney	 & 223.06-242.93 & 2017 July 09-14:43 &	 40	&  16-2647	 &	12  &		DA62 	  \\
2015.1.00118.S & L.~Looney	 & 223.06-242.93 & 2017 July 09-15:53 &  40	&  16-2647	 &	12  &		DA62 	  \\	
2018.1.00945.S & C.~Qi		 & 217.06-235.43 & 2018 Oct  26-04:53 &	 49	&  15-1397	 &	27  &		DA43  \\	
2018.1.00945.S & C.~Qi		 & 217.06-235.43 & 2018 Nov  17-03:55 &	 46	&  15-1397	 &	27  &		DA43	  \\		
2018.1.00945.S & C.~Qi		 & 217.06-235.43 & 2019 July 07-12:55 &  45	&  149-13894 &	29  &		DV09  \\	
2018.1.00945.S & C.~Qi		 & 217.06-235.43 & 2019 July 07-13:56 &  45	&  149-13894 &	29  &		DV09	  \\			
2018.1.01255.S & M.~Benisty	 & 213.06-229.93 & 2019 July 13-15:23 &	 40	&  111-12644 &	35  &		DV09  \\	
2018.1.01255.S & M.~Benisty	 & 213.06-229.93 & 2019 July 19-14:48 &	 43	&  95-8547	 &	35  &		DV09	  \\
2018.1.01255.S & M.~Benisty	 & 213.07-229.94 & 2021 Aug  20-11:28 &	 45	&  59-9933	 &	35  &		DA60  \\
\hline
\multicolumn{8}{c}{Band 7 Observations} \\ 
\hline
2012.1.00870.S & L.~Perez	 & 339.09-342.96 & 2014 Aug  29-09:44 &  36	&  33-1090	 &	35  & 		DA63  \\			
2018.1.00350.S & A.~Isella	 & 332.86-348.74 & 2019 July 21-11:23 &	 44	&  92-8547	 &	39  &		DV05  \\		
2018.1.00350.S & A.~Isella	 & 332.86-348.74 & 2019 July 28-11:06 &	 42	&  92-8547	 &	39  &		DV05  \\
2018.1.00350.S & A.~Isella	 & 332.86-348.74 & 2019 July 30-10:21 &  46	&  92-8547	 &	39  &		DA52  \\
2018.1.00350.S & A.~Isella	 & 332.86-348.74 & 2019 July 30-11:51 &	 46	&  92-8547	 &	39  &		DA52  \\
\enddata
\tablecomments{Col.~(1): ALMA program ID. Col.~(2): PI of the ALMA project. Col.~(3): Frequency range used in the combined dataset. Col.~(4): UTC time at the start of the observation. Col.~(5): Number of antennas. Col.~(6): Minimum and maximum baseline lengths. Col.~(7): On-source observing time. Col.~(8): Reference antenna used in data calibration. }
\end{deluxetable*}

\section{Observations and Data Reduction} \label{sec:obs}

%%%%%%%%%%%%%%%%%%%%%%%%%%%%%%%%%
%%%%%%%%%%%%%%%%%%%%%%%%%%%%%%%%%
\begin{figure*}[!t]
\centering
    \includegraphics[width=\textwidth]{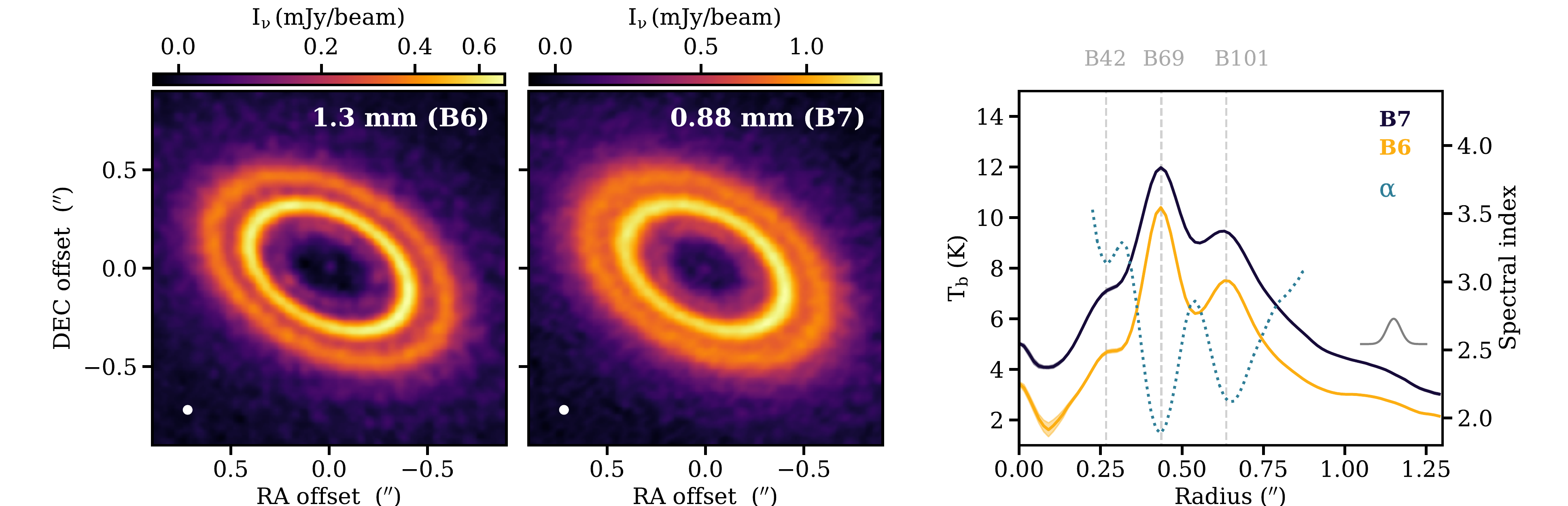}
\caption{{\bf \textit{Left and Middle:}} Continuum emission images of the LkCa~15 disk at 1.3 and 0.88\,mm, respectively, with identical beam size of 50\,mas, shown in the bottom-left corner of each panel. An asinh color stretch is applied to highlight the faint emission.
{\bf \textit{Right:}} Azimuthally-averaged radial intensity profiles in brightness temperature. The three prominent dust rings are marked by the dashed vertical lines. The spectral index profile between the two bands is indicated by the dotted line ($y-$axis labels on the right side). The Gaussian profile in the bottom-right corner shows the FWHM of the synthesized beam. 
  \label{fig:cont-maps}}
\end{figure*}
%%%%%%%%%%%%%%%%%%%%%%%%%%%%%%%%%
%%%%%%%%%%%%%%%%%%%%%%%%%%%%%%%%%

Our analysis included ALMA continuum data with both the Band 6 (B6) and Band 7 (B7) receivers. A summary of the observations and instrument configurations is provided in Table~\ref{tab:obs}.
For individual executions, calibration scripts provided by the observatory were used with the specified {\tt CASA} versions \citep{McMullin2007} to flag problematic data, correct for bandpass responses, set absolute flux scales, and solve for complex gain variations. Self-calibration and imaging were then performed with \texttt{CASA 6.4.0}. Before concatenating data taken at different epochs, we shifted all measurement sets to a common phase center. To do so, we first created a continuum image for each data set and then determined the disk emission center in the image plane. For short-baseline data, the disk center was estimated by fitting a 2D Gaussian model with \texttt{imfit}. For long-baseline data, we applied an ellipse model to the prominent dust ring at 69\,au (see Section~\ref{sec:results}) and adopted its centroid as the disk center. The \texttt{phasevis} and \texttt{fixplanets} tasks in \texttt{CASA} were then used to implement these shifts.

Self-calibration was done separately for each band, starting from the concatenated short-baseline data sets. The initial model was created with the \texttt{multiscale} algorithm implemented in the \texttt{tclean} task and a Briggs weighting parameter of 0.5. After two rounds of phase (solutions on scan-length and 60\,s) and one round of amplitude self-calibration (on a scan-length interval), the peak S/N in the image was nearly doubled. The improvement for image quality was minimal for self-calibrating the combined data (including long baselines). One subsequent iteration of phase self-calibration was performed with a solution interval of 900\,s and spectral windows combined.

In synthesizing a B6 continuum image, we found a good compromise between resolution and sensitivity was reached with \texttt{robust=0} in the Briggs weighting scheme, resulting in a beam of 47$\times$34\,mas (PA=$-$3.3$\degr$). %and rms noise level of 9.6\,$\mu$Jy beam$^{-1}$. 
As discussed in \citet{JvM1995}, the mismatch in units between CLEAN residual map (Jy per dirty beam) and the convolved CLEAN model image (Jy per clean beam) would lead to inaccurate source flux and mis-representation of faint emission. Following the procedure outlined in \citet{Czekala2021}, we rescaled the clean image residuals by the ratio of the clean and dirty beam areas ($\epsilon$=0.45) before adding to the convolved model. This effectively reduces the image noise level to 4.4\,$\mu$Jy/beam\footnote{The threshold in \texttt{tclean} was set to 25 and 50\,$\mu$Jy in B6 and B7, respectively. Compared to \citet{Facchini2020}, the combined data in this study are more sensitive by a factor of $\sim$1.5 and reach a better angular resolution by 30\%.}. We note that the main results are not affected by this correction and the detection significance from the original image is discussed accordingly. 

A high-resolution B7 continuum image with a beam size of 46$\times$38\,mas (PA=51.3$\degr$) was produced with Briggs weighting of \texttt{robust=0} and a Gaussian uvtaper (10$\times$40\,mas, 0$\degr$) using \texttt{tclean}. After correcting for the unit mismatch with $\epsilon$=0.72, the rms noise level for the B7 continuum image was 15.5\,$\mu$Jy/beam. 
As a last step, we convolved the final images with an elliptical Gaussian kernel so they both have a circular beam with a full width at half maximum of 50\,mas (8\,au).  This ensures a fair comparison between the two wavelengths.

\section{The Dust Disk of LkCa~15} \label{sec:results}
% overall description of the disk emission 
% inner disk + three rings + outer tail 

\subsection{Overall Dust Emission Morphology in LkCa~15}
Figure~\ref{fig:cont-maps} presents the new deep continuum images of the LkCa~15 disk at both 1.3\,mm (B6) and 0.88\,mm (B7), as well as the corresponding azimuthally-averaged radial profiles, deprojected with an inclination angle of 50\fdg2 and position angle of 61\fdg9 (counting from N to E,  \citealt{Facchini2020}). The emission morphologies at the two wavelengths are very similar, though the emission at 1.3\,mm exhibits higher contrasts. 
%at 0.88\,mm generally appears more diffuse. 
The disk shows a faint inner component, a wide cavity, three dust rings spanning 0\farcs2--0\farcs8, and an outer tail extending beyond $\sim$1\arcsec. \texttt{CASA imfit} suggests the inner disk component is unresolved ($<$4\,au in radius), with a spectral index of 2.4$\pm$0.3 based on peak intensities of $58.6\pm4.1$ and $155.8\pm7.9\,\mu$Jy/beam at B6 and B7, respectively.
Around the radial distances of the three proposed giant planet candidates (0\farcs10--0\farcs13; \citealt{Kraus2012, Sallum2015}), we do not find any significant continuum emission.

There are three dust rings peaking at 42, 69, and 101\,au (hereafter B42, B69, and B101), respectively, with widths comparable to the beam size (based on model fitting described below). 
The B42 ring corresponds to the emission ``shoulder" reported in \citet{Facchini2020}, which is better resolved in the data presented here. The two bright dust rings B69 and B101 appear to be narrower (by $\gtrsim$20\% in width, see Table~\ref{tab:frank} in the Appendix) at 1.3\,mm compared to 0.88\,mm, which suggests that dust grains are likely trapped in local pressure bumps with larger grains being more effectively concentrated \citep{Pinilla2015, Long2020}. 
We obtained a spectral index map from the images at the two bands as $\alpha_{\rm mm}={\rm log}(I_{B6}/I_{B7})/{\rm log}(\nu_{B6}/\nu_{B7})$.
The radial profile of $\alpha_{\rm mm}$ (the right panel of Figure~\ref{fig:cont-maps}) reaches local minima at ring peaks and maxima in gaps. Inferring the grain properties in B69 and B101 rings is challenging because the dust emission is likely optically thick (with low $\alpha_{\rm mm}\sim2$).
The overall higher $\alpha_{\rm mm}$ around B42 however suggests that grain properties there might be different from the outer bright disk regions.

\subsection{Asymmetric Features in the B42 Ring} \label{sec:asymmetric}
\begin{figure}[!t]
\centering
    \includegraphics[width=0.4\textwidth]{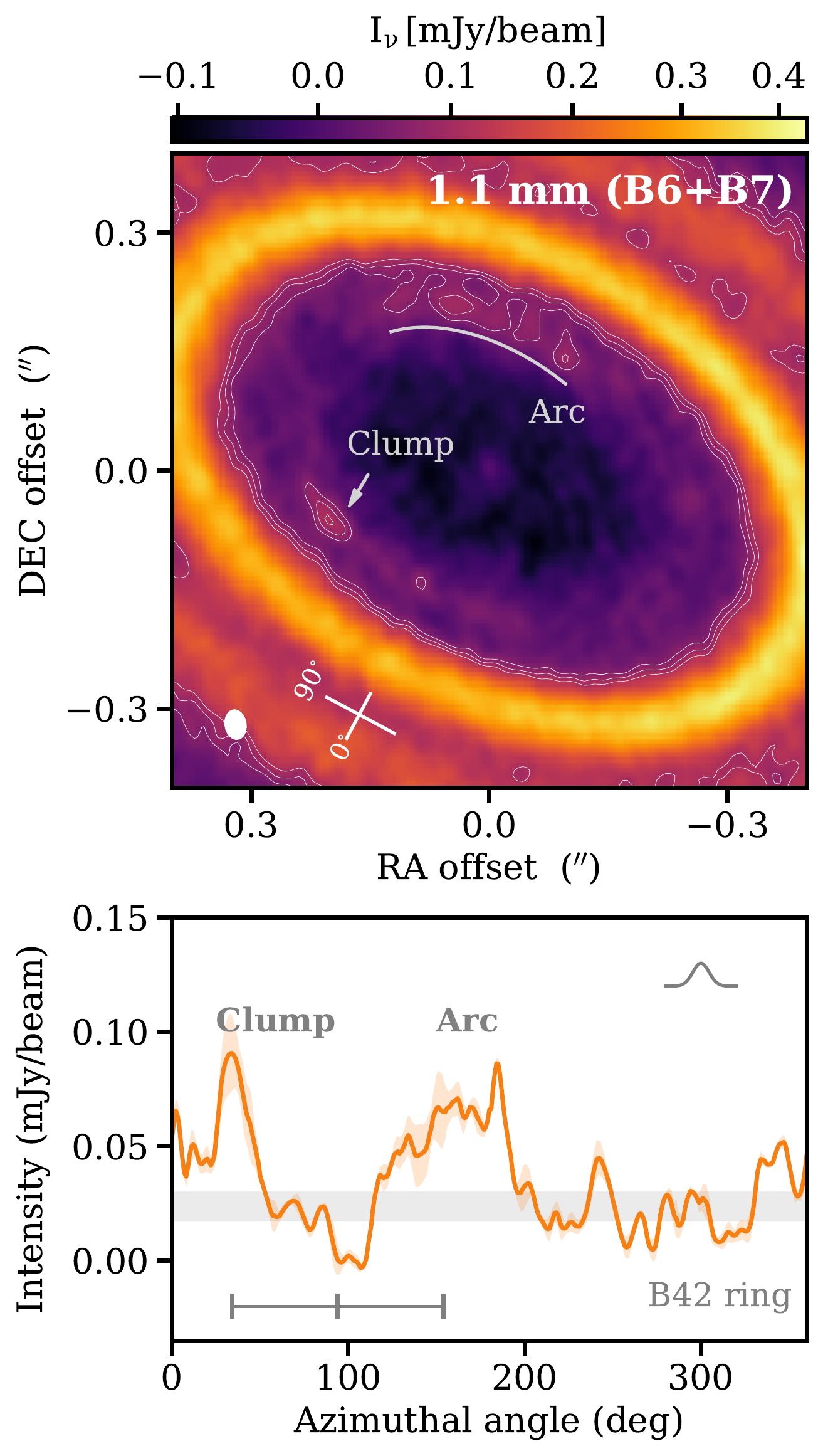}
\caption{{\bf \textit{Top:}} Continuum image of the combination of 1.3 and 0.88\,mm data (a zoom-in view), with contours at 6, 8, 10$\sigma$. The insert cross in the bottom-left corner marks the azimuthal angle convention used in this work. {\bf \textit{Bottom:}} The azimuthal intensity profile along the B42 ring, obtained as the average emission from $0\farcs26-0\farcs28$ in the disk plane. The clump peak is at 32$\degr$ and one grey tick indicates 60$\degr$ separation. The grey shaded region indicates the standard deviation of emission within 200$\degr-360\degr$. The Gaussian profile in the upper right corner shows the FWHM of the synthesized beam.\label{fig:2band}}
\end{figure}

%Excess Emission in the Faint Dust Ring
The innermost faint ring, B42, exhibits azimuthal asymmetries, which are better visualized in the two-wavelength combined image (at 1.1\,mm) in Figure~\ref{fig:2band}. That image was constructed with \texttt{robust=-0.2} in \texttt{tclean} using the \texttt{mtmfs} algorithm \citep{RauCornwell2011} and \texttt{nterms=2} to obtain a high angular resolution of 39$\times$28 mas. Using an azimuthal intensity profile around this ring (made with a radial average from $0\farcs26-0\farcs28$ in the disk plane), we identified two prominent features that are separated by $\sim$120\degr\ (see bottom panel of Figure~\ref{fig:2band}). We refer to these features as ``clump'' and ``arc'', based on the apparent differences in their azimuthal extensions. The clump and arc are brighter than the average of the B42 ring by about a factor of two.

%%%%%%%%%%%%%%%%%%%%%%%%%%%%%%%%%
%%%%%%%%%%%%%%%%%%%%%%%%%%%%%%%%%
%%%%%%%%%%%%%%%%%%%%%%%%%%%%%%%%%
%%%%%%%%%%%%%%%%%%%%%%%%%%%%%%%%%

%%%%%%%%%%%%%%%%%%%%%%%%%%%%%%%%%
%%%%%%%%%%%%%%%%%%%%%%%%%%%%%%%%%
\begin{figure*}[!t]
\centering
    \includegraphics[width=0.8\textwidth, trim=10 0 0 0 , clip]{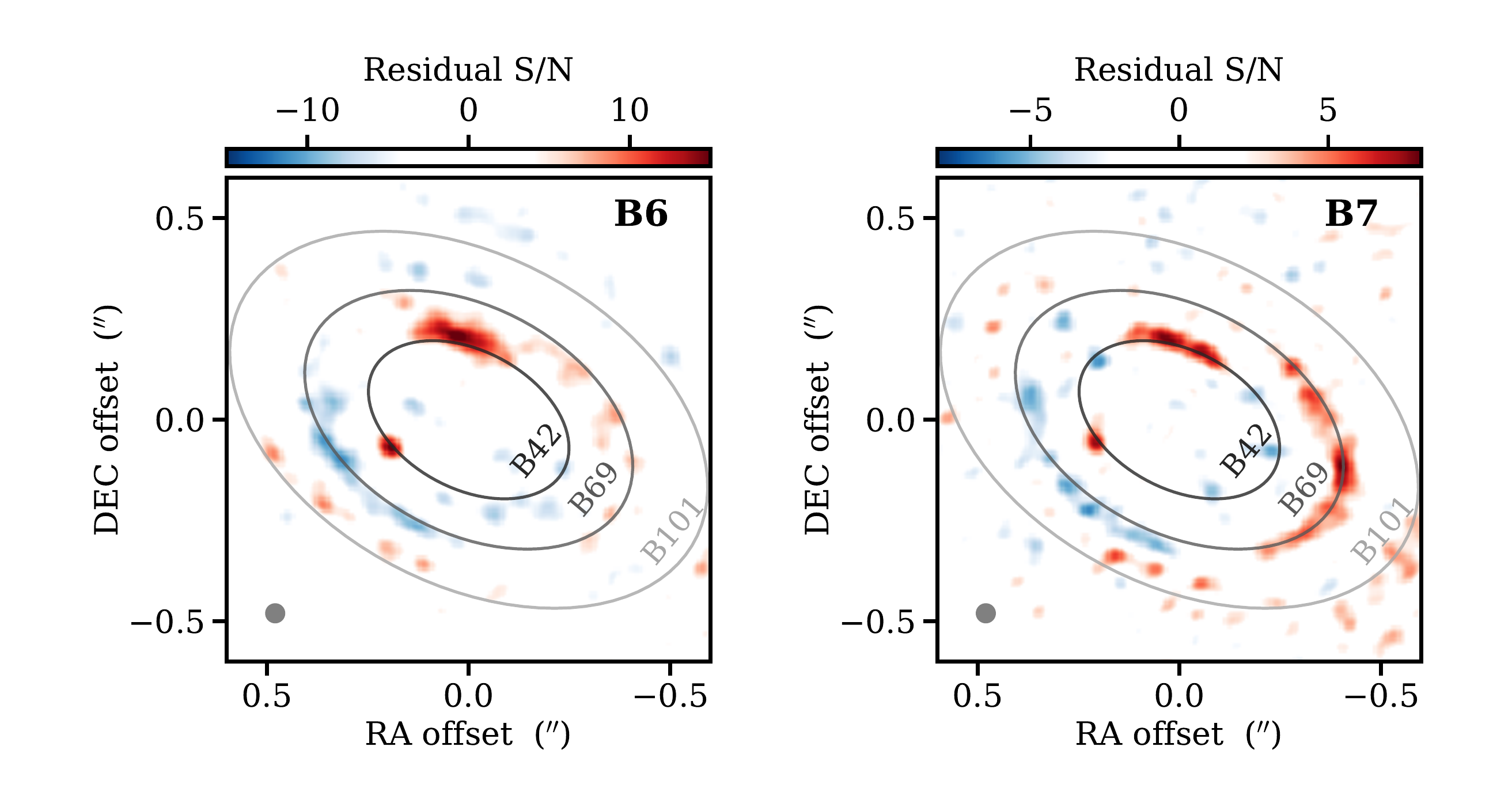} \\
    \includegraphics[width=0.8\textwidth, trim=10 0 0 85 , clip]{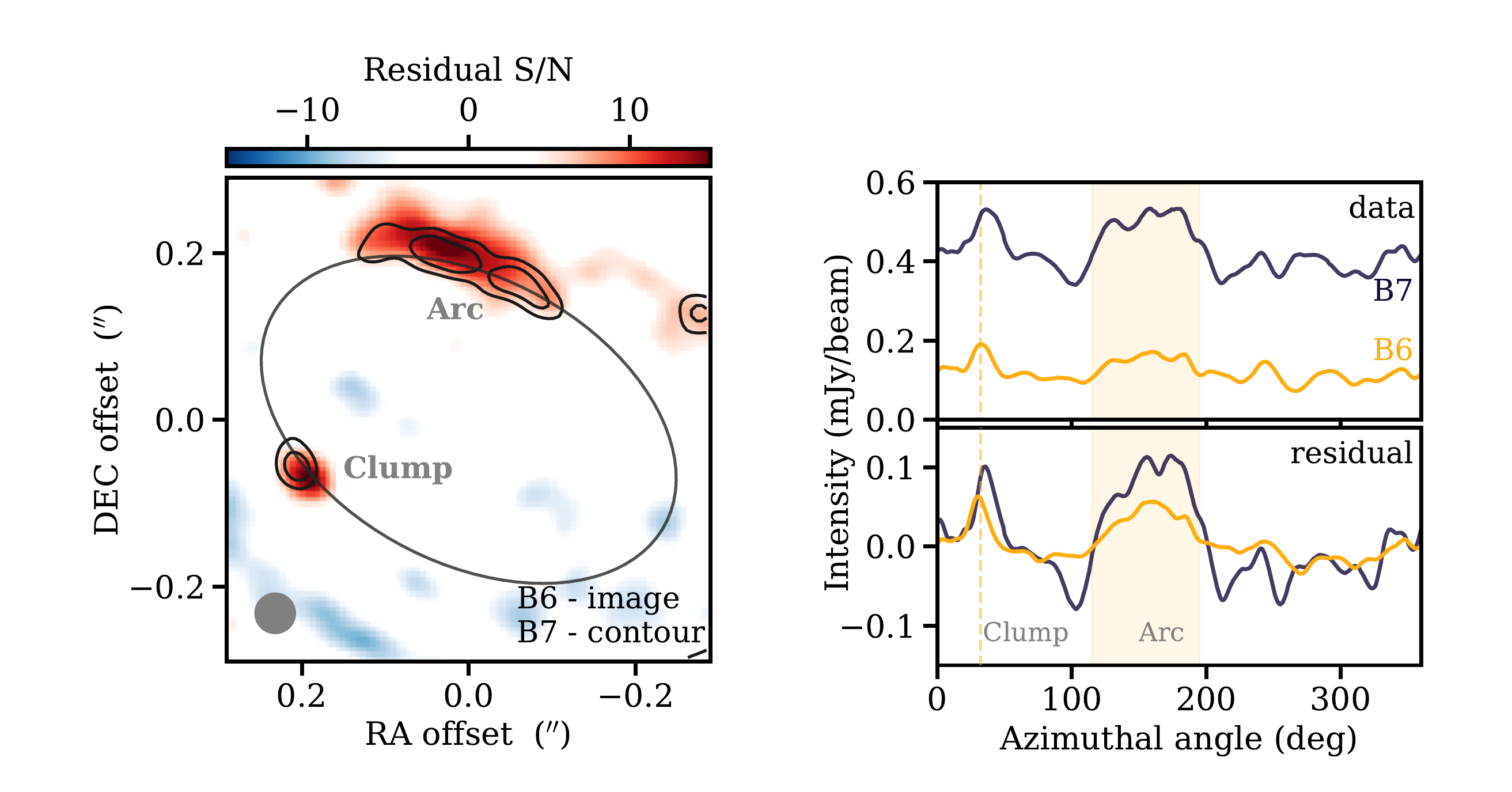} \\
\caption{{\bf \textit{Upper Panels:}} Residual images after subtracting the \texttt{frank} model, shown in signal-to-noise ratio. The images are convolved with the same circular beam of 50\,mas, with noise levels of 4.4 and 15.5$\mu$Jy/beam at B6 and B7, respectively. The three ellipses indicate the derived ring locations; {\bf \textit{Lower Left:}} The zoom-in comparison of the two residual images for the relative locations of the two excess emission features, with B7 residuals shown in contours at 4 and 6$\sigma$;  {\bf \textit{Lower Right:}} The B42 azimuthal intensity profiles from the data (upper) and residual images (lower). The vertical dashed line marks the peak location of the clump identified in the B6 data at 32$\degr$ and the shaded region is a rough estimate of the arc extension from 115--195$\degr$. \label{fig:2band-res}}
\end{figure*}

%%%%%%%%%%%%%%%%%%%%%%%%%%%%%%%%%
%%%%%%%%%%%%%%%%%%%%%%%%%%%%%%%%%

To better characterize these nonaxisymmetric features, we first built an azimuthally symmetric disk model and then analyzed a residual map where that symmetric emission was removed. This was achieved by fitting the real component of the visibilities with the \texttt{frank} software package \citep{Jennings2020} and subtracting the model from the data to create a set of residual visibilities, from which a residual map was then produced with \texttt{tclean} (hereafter referred as `residual' throughout the paper). This procedure has been well established in previous studies \citep[e.g.,][]{Andrews2021,Jennings2022}. 
We set the \texttt{frank} hyperparameters to $R_{\rm out}=3''$ (to encompass all the emission), $\alpha=1.4$ and $\omega_{\rm smooth}=0.1$ (to damp high-frequency oscillations in the model brightness profile). We fixed the disk inclination of 50\fdg2 and position angle to 61\degr9 \citep{Facchini2020}, and determined phase-center offsets by fitting the ellipse centroid of the bright B69 ring in the image.  We estimated $\delta_{\rm RA}=0.4\,$mas and $\delta_{\rm Dec}=0.2\,$mas for B6, and $\delta_{\rm RA}=-1.5\,$mas and $\delta_{\rm Dec}=-1.0\,$mas for B7.

The images synthesized from the residual visibilities are shown in Figure~\ref{fig:2band-res} (the full \texttt{frank} fitting results, including the comparison of data and model, are presented in Appendix~\ref{sec:frank}). The two non-axisymmetric emission features are clearly seen along the B42 ring at both wavelengths, at significance levels of 15$\sigma$ and 8$\sigma$ for B6 and B7, respectively (the 1$\sigma$ noise is measured as the rms scatter in an annulus from 4 to 8\arcsec\ from the disk center). Considering the images without JvM correction, the excess emission features are detected at 10$\sigma$ and 6$\sigma$ for B6 and B7, respectively.
Evident residuals are also visible at larger disk radii, which could be due to elevated emission surfaces, mismatches in the geometry properties of individual rings, or perturbations from embedded planets (see also Figure~\ref{fig:model}).
%or their combinations.
Varying the inclination and position angles by 1$\degr$ or the center offsets by 3\,mas (a half pixel in the image) from the default values produces even more prominent residuals,  in particular on large scales. Regardless, in all cases the two asymmetric features in the B42 ring are preserved and stand out. Note that significance of these emission features is evaluated based on rms noise defined in emission-free regions at large scales. If we define the noise level in local regions along the ring (in azimuths $>200\degr$), the clump and arc are also clearly detected at both bands (4--5$\sigma$).

The clump is unresolved in the residual maps. It has a flux density of $\sim$0.08 and 0.13\,mJy at 1.3 and 0.88\,mm, respectively, corresponding to a dust mass of 0.03--0.07\,$M_{\oplus}$ for a dust temperature of 20\,K and the DSHARP opacity with a maximum grain size of 1\,mm ($\kappa_{\rm 1.3\,mm}$=1.9\,cm$^2$\,g\,$^{-1}$ and $\kappa_{\rm 0.88\,mm}$=3.6\,cm$^2$\,g\,$^{-1}$, \citealt{Birnstiel2018}).
The arc is resolved in 3--4 independent beams along the ring in the azimuthal direction, with integrated fluxes about 4 times higher than the clump.

When comparing the non-axisymmetric features at the two wavelengths, subtle differences in their locations were identified. The peak position of the B7 clump is offset by 5$\degr$ clockwise from B6. 
The offset estimated from the data directly (before model subtraction) is slightly smaller (offset of 4$\degr$; see the lower right panel of Figure~\ref{fig:2band-res}).
This difference can be explained by the fluctuations (or intrinsic asymmetry) along the underlying dust ring. For example, brighter emission is seen to the southwest (smaller azimuthal angle) of the clump in B6 that can cause the shift of the clump peak in the residual map. To test if the offset between the two bands can be caused by CLEAN artifacts, we created mock visibilities by injecting a point source with comparable flux to the clump along the B42 ring in the data (500 samples with randomly assigned azimuthal locations, excluding regions of the excess emission features) and estimated its azimuthal shift in the recovered residual image. In most cases ($\sim80\%$), the shift is within $1\degr$. 
We also explored if the shift could be due to orbital motion. We made images with only the long-baseline executions conducted in 2019 July, and similar offsets were observed. 
Therefore, this peak shift might reflect a truly different spatial distribution of grains with different sizes. At a semimajor axis of 42\,au, a full orbit takes 246.9\,yr and the clump will move by $\sim1.5\degr/\rm yr$ anticlockwise if following Keplerian motion. Such movement could be witnessed with new observations taken from now on, given a minimum three-year time baseline.

We assigned an azimuthal range of 115--195$\degr$ for the arc feature (the azimuthal angle convention is shown in Figure~\ref{fig:2band}), though the edges of this feature are hard to determine. While the B6 residual arc appears more centrally peaked and the B7 one appears more clumpy, quantifying any spatial variations between the two wavelengths is challenging given the low surface brightnesses. Nevertheless, the main body of the arc seems to reside outside the peak of the B42 ring. Possible explanations include a slightly eccentric inner ring and elevated emission surface of the arc.

\section{Discussion} \label{sec:diss}
We have detected non-axisymmetric dust continuum emission features with a (point-like) clump and (more extended) arc morphology separated by $\sim$120\degr\ in azimuth along the faint B42 ring in the LkCa 15 disk.  These features are robust, as they are seen with high S/N in independent observations at two different wavelengths. 
Though the flux of the clump is comparable to predictions of CPD emission levels for Jupiter-mass planets \citep{Zhu2018}, the observed azimuthal shift between the B6 and B7 images that could not be accounted for by Keplerian motion makes the possibility that it represents a CPD unlikely.
Instead, the fact that the clump and the arc are separated by roughly 120$\degr$ suggests that the B42 ring may trace dust particles co-orbiting with an embedded planet (in ``horseshoe'' orbits) with enhanced accumulation in the vicinity of the $L_{4}$ and $L_{5}$ Lagrangian points.

%%%%%%%%%%%%%%%%%%%%%%%%%%%%%%%%%
%%%%%%%%%%%%%%%%%%%%%%%%%%%%%%%%%
\begin{figure}[!t]
\centering
    \includegraphics[width=0.5\textwidth, trim=20 80 0 80, clip]{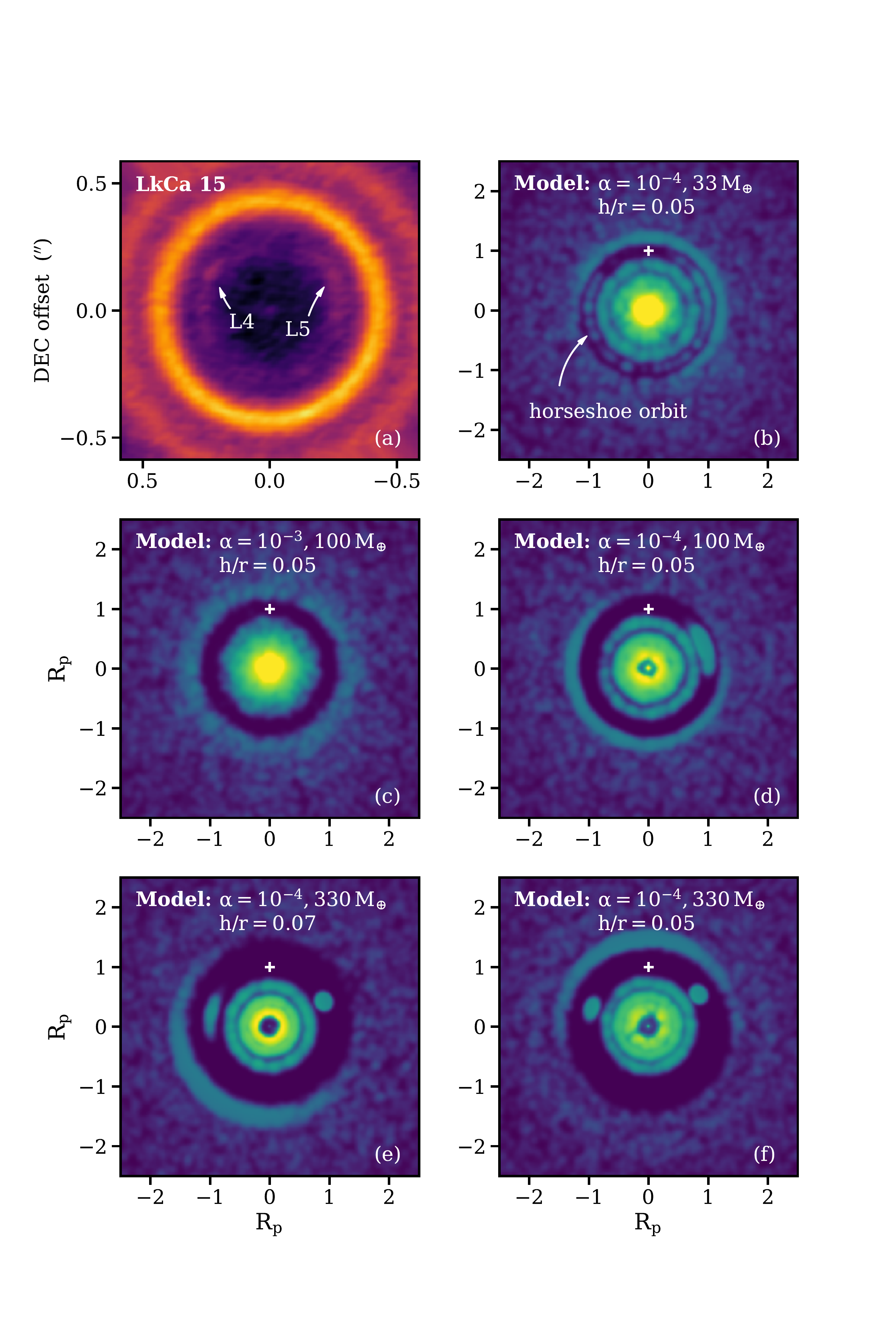} 
\caption{{\bf \textit{Panel (a):}} The deprojected  1.1 mm (B6+B7 combined) image of the LkCa~15 disk, with the $L_{4}$ and $L_{5}$ points marked. {\bf \textit{Panels (b-f):}} 1.3\,mm continuum images from planet--disk interaction models selected from \citet{Zhang2018}, assuming the maximum grain size of 1\,mm. In each panel, the planet is located at (0, 1) and marked with a `+' symbol; it orbits in the counterclockwise direction, following the $L_{4}$/$L_{5}$ annotation in panel (a). 
 \label{fig:model}}
\end{figure}
%%%%%%%%%%%%%%%%%%%%%%%%%%%%%%%%%
%%%%%%%%%%%%%%%%%%%%%%%%%%%%%%%%%

In the framework of planet--disk interactions, the propagation of density waves creates gaps on both sides of the planet where dust particles are trapped at the gap edges, while a fraction of dust co-moves with the planet in a horseshoe-shaped region \citep[e.g.,][]{Kley2012, Paardekooper2022_ppvii}. The dust horseshoe ring has been explored in hydrodynamical simulations (e.g., \citealt{Dong2017_superearth,Ricci2018}), and has been advocated as an explanation for certain ALMA observations (e.g., HL Tau, \citealt{Dipierro2015}; HD 169142, \citealt{Perez2019}). The morphology of the horseshoe ring depends on many disk and planet properties, including the degree of dust--gas coupling, disk viscosity, local thermodynamics (i.e., scale height aspect ratio $h/r$), and planet mass, among others. 
Figure~\ref{fig:model} illustrates some of the key dependencies using a subset of the 2D hydrodynamical models \footnote{CPDs were not included in these simulations, as dust evolution in CPDs has not been well understood. Given the small scale of a CPD, its influence on dust evolution in the horseshoe region should be minimal.}
calculated by \citet{Zhang2018}. 
Higher viscosity (characterized by the $\alpha$ parameter, Panels c/d) tends to destroy the horseshoe ring, as diffusion smooths out the gas density perturbations (see also \citealt{Rodenkirch2021}). 
More massive planets generate stronger non-axisymmetric features, with higher dust concentration around the Lagrangian points (Panels b/d/f).

Our observations of the B42 dust ring in the LkCa~15 disk best resemble the scenario where a Neptune-mass planet is perturbing a low-viscosity disk (Panels a/b in Figure~\ref{fig:model}). Based on the gas emission morphology (\citealt{Oberg2010}, \citealt{Jin2019} separating front and back disk sides), the disk around LkCa~15 rotates counterclockwise. Therefore, the clump and arc features are likely marking dust associated with the leading ($L_4$) and trailing ($L_5$) Lagrangian points, respectively. 

\citet{Montesinos2020} explored the evolution of dust around Lagrangian points when the disk is shaped by interactions with a planet and suggested that $L_4$ and $L_5$ can act as dust traps. 
However, they found the dust concentration at Lagrangian points is highly dynamic, with morphologies quickly evolving with time. In addition, multiple physical processes (e.g., drag forces, grain collisions and growth) could modify these dust distributions, resulting in different emission morphologies (e.g., dust feedback can lead to fragmentation of the dust clump, \citealt{Rodenkirch2021}). Therefore, the clump and arc morphologies in the B42 ring may be transient: using them to infer the planet and disk properties is not trivial.

The shift of the clump peak that was discussed in Section~\ref{sec:asymmetric} was also predicted by \citet{Rodenkirch2021}, when modeling the HD\,163296 disk. They found that smaller grains with lower Stokes number around $L_4$ are located closer to the planet (see their Figures~8 and 9). The brighter emission in the arc is also consistent with their prediction that larger grains are preferentially trapped at the $L_5$ point. The accumulation of dust particles in the vicinity of Lagrangian points may trigger the formation of Trojan asteroids or even Earth-mass bodies around massive planets (e.g., \citealt{Lyra2009}). The estimated dust mass in the clump and arc is sufficient to account for the Jovian Trojans in our Solar System, though slightly more Trojans are found around $L_4$ \citep{Vinogradova2015SoSy}, in contrast with our inference of more dust mass in $L_5$.  

% The total mass of the Jupiter trojans is estimated at 0.0001 of the mass of Earth or one-fifth of the mass of the asteroid belt 

%%%%%%%%%%%%%%%%%%%%%%%%%%%%%%%%%
%%%%%%%%%%%%%%%%%%%%%%%%%%%%%%%%%
\begin{figure}[!t]
\centering
    \includegraphics[width=0.45\textwidth]{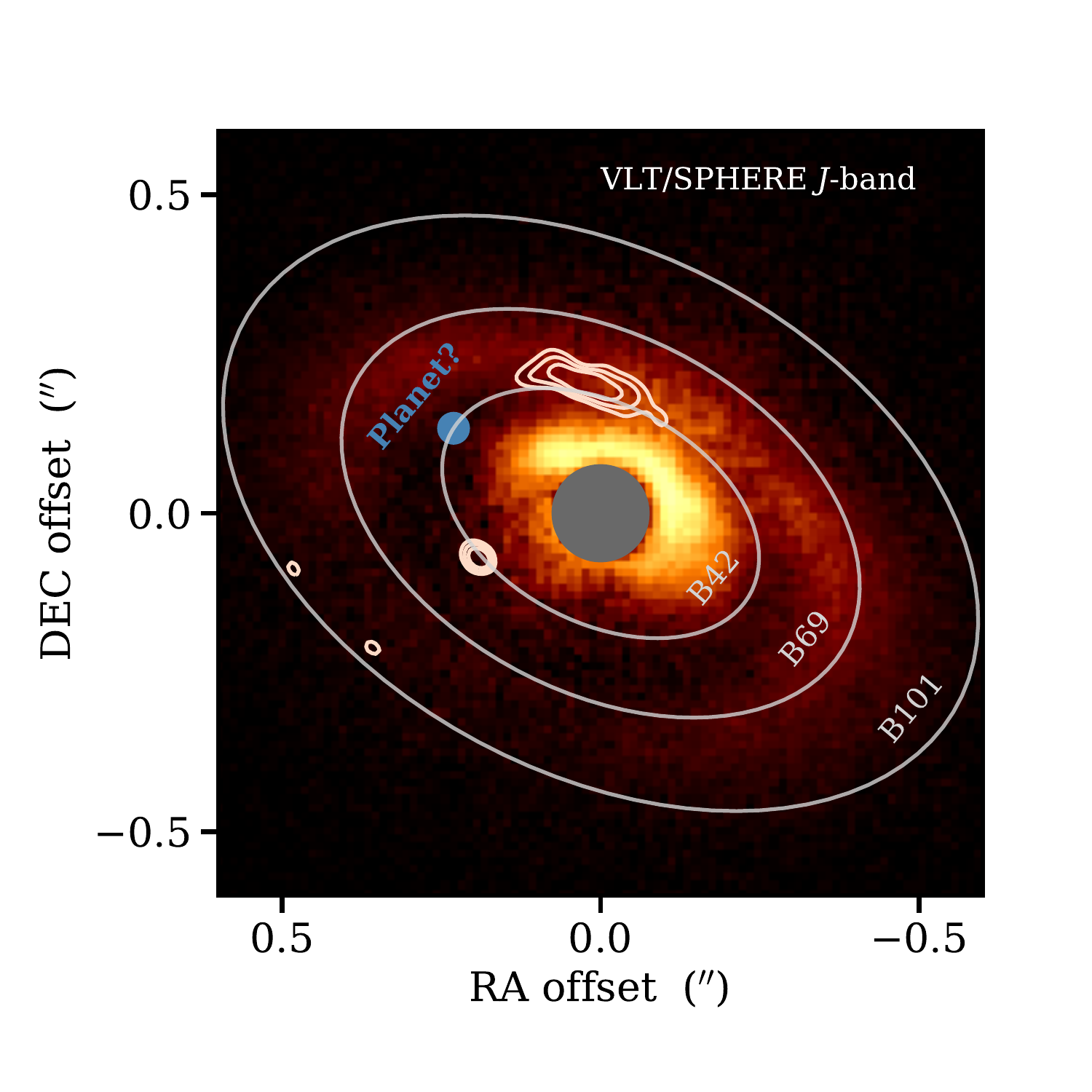}
\caption{The comparison of near-IR dust scattered light (SPHERE IRDIS $J$-band imaging polarimeter, \citealt{Thalmann2016}) and ALMA millimeter dust rings (the three grey ellipses) in the LkCa~15 disk. The inner working angle of 80\,mas in the SPHERE image is masked out. The ALMA B6 residual contours are shown at 8, 10, 12$\sigma$ levels in pink. The putative planet location is indicated. \label{fig:sphere}}
\end{figure}
%%%%%%%%%%%%%%%%%%%%%%%%%%%%%%%%%
%%%%%%%%%%%%%%%%%%%%%%%%%%%%%%%%%

Based on the identification of the clump ($L_4$) and the arc ($L_5$), we can infer that the putative planet location along the B42 ring is roughly aligned with the major axis of the LkCa~15 disk to the northeast as indicated in Figure~\ref{fig:sphere}. Along the ring, it is interesting that the vicinity of the planet is comparatively depleted of dust emission in the ALMA data (see Figure~\ref{fig:2band}). 
As we are interested in localized emission within the horseshoe ring, we estimate the noise level from the standard deviation along the B42 ring (from azimuths $>$200$\degr$) and find a 3$\sigma$ upper limit of $\sim$70$\,\mu$Jy at 0.88\,mm ($\sim$33$\,\mu$Jy at 1.3\,mm) for any associated CPD, which translates into a dust mass upper limit of 0.02-0.03\,$M_{\oplus}$ following the same assumptions as in Section~\ref{sec:asymmetric}.
This is a bit higher than the emission levels seen from the planet PDS\,70\,c (46$\,\mu$Jy; \citealt{Benisty2021}) and planetary-mass object SR\,12\,c (63$\,\mu$Jy; \citealt{Wu2022}) at 0.88\,mm, when normalized to the LkCa\,15 distance. Deeper ALMA observations are needed to put comparable constraints on the presence of a CPD in the LkCa\,15 disk.

For disks with inner cavities, the masses of perturbers are hard to constrain because the gap width opened by each perturber is largely unknown. In the case of LkCa\,15, gap width associated with the B42 planet can be estimated based on locations of the horseshoe ring and the adjacent outer ring (B69), following the definition in \citet{Zhang2018}. Assuming canonical disk conditions of $h/r=0.05$, $\alpha=10^{-4}$ and maximum grain size of 1\,mm, a fractional gap width of 0.52 corresponds to a planet mass of 0.1--0.3\,$M_{\rm Jup}$ (Figure~12 in \citealt{Zhang2018}, see also \citealt{Bae2017}), which is below the current detection limit from direct imaging for LkCa\,15 at 42\,au \citep{Asensio-Torres2021}.

Additional constraints on the planet properties can be made by comparing different disk tracers. 
Interestingly, the gap seen in the near-IR scattered light image \citep{Thalmann2016} shown in Figure~\ref{fig:sphere} is coincident with the planet horseshoe ring identified from the millimeter dust emission. The absence of scattered light is likely due to inner disk shadowing for this low dust surface density area. 
Based on the gap morphology in the scattered light, \citet{DongFung2017} predicted a planet mass of 0.15\,$M_{\rm Jup}$ (1.5\,$M_{\rm Jup}$) for $\alpha=10^{-4}$ ($\alpha=10^{-2}$) at 42\,au with hydrodynamic simulations. The presence of material along the horseshoe orbit from the ALMA data presented here would prefer the lower planet mass scenario. As a higher mass planet would more effectively prevent gas flowing through the cavity, this low mass scenario is also more consistent with the presence of substantial amounts of molecular gas in the inner LkCa~15 disk \citep{Leemker2022arXiv}. Nevertheless, this B42 planet may not be sufficient to fully account for both the wide inner cavity and the B101 ring further out, where additional planets and/or other mechanisms might act in concert to shape the LkCa\,15 disk.

Though we favor this Lagrangian point dust trapping scenario for dust emission around 42\,au as it explains the clump and arc features simultaneously, there are also other possibilities. The arc could be part of a spiral arm (e.g., \citealt{Andrews2021}), which might explain its radial offset from the ring center. Both the arc and clump could be vortex, for which gas kinematics that traces the anticyclonic motion will help identifying their nature \citep{Boehler2021}. 
Meanwhile, searching for gas kinematic perturbations near the predicted location of the B42 planet would be especially compelling to test our proposed scenario.

\section{Summary} \label{sec:summ}

In this Letter, we reported a detailed investigation of high quality ALMA continuum observations of newly-identified dust features in the LkCa 15 disk.  We found that the continuum emission ring at a radius of 42 au looks very similar to the horseshoe ring produced in planet--disk interaction models. We postulated that the two robustly detected non-axisymmetric emission features along this ring with a separation of $\sim$120$\degr$ mark dust concentrations at the $L_{4}$ and $L_{5}$ Lagrangian points, which are consistent with hydrodynamic model predictions for interactions with an embedded $\sim$Neptune-mass planet.  Multiple additional lines of evidence, including dust trapping in a ring at 69 au and the fact that the 42 au ring coincides with a gap in the near-IR scattered light also support the scenario of an embedded planet.  
This putative planet should be located at a projected separation of $0\farcs27$ from the star with a position angle of $\sim60\degr$ (roughly aligned with the disk major axis). 
However, no point-like CPD emission is identified, with a 3$\sigma$ upper limit of $\sim$70$\mu$Jy at 0.88\,mm ($\sim$33$\mu$Jy at 1.3\,mm) along the ring. 

These results demonstrate the utility of deep high-resolution ALMA images in revealing the faint, subtle features predicted by hydrodynamic simulations. These features provide compelling new evidence of ongoing planet formation in structured disks and offer a new avenue for guiding planet searches.

\paragraph{Acknowledgments}
We thank the referees for their helpful comments that improved the clarity of this manuscript. F.L. is grateful to Gregory Herczeg, Rixin Li, and Yifan Zhou for support. 
F.L. acknowledge support from the Smithsonian Institution as the Submillimeter Array (SMA) Fellow. F.L. and S.A. acknowledge funding support from the National Aeronautics and Space Administration under grant No.17-XRP17$\_$2-0012 issued through the Exoplanets Research Program. A.I. acknowledges support from the National Aeronautics and Space Administration under grant No. 80NSSC18K0828. Support for J.H. was provided by NASA through the NASA Hubble Fellowship grant \#HST-HF2-51460.001-A awarded by the Space Telescope Science Institute, which is operated by the Association of Universities for Research in Astronomy, Inc., for NASA, under contract
NAS5-26555. S.Z. and Z.Z. acknowledge support from NASA through the NASA FINNEST grant 80NSSC20K1376. Z.Z. acknowledges support from the National Science Foundation under CAREER grant AST-1753168.  This project has received funding from the European Research Council (ERC) under the European Union’s Horizon 2020 research and innovation programme (PROTOPLANETS, grant agreement No. 101002188).

This paper makes use of the following ALMA data sets: 2012.1.00870.S, 2015.1.00118.S, 2018.1.00945.S, 2018.1.01255.S, and 2018.1.00350.S. ALMA is a partnership of ESO (representing its member states), NSF (USA), and NINS (Japan), together with NRC (Canada), MOST and ASIAA (Taiwan), and KASI (Republic of Korea), in cooperation with the Republic of Chile. The Joint ALMA Observatory is operated by ESO, AUI/NRAO, and NAOJ. The National Radio Astronomy Observatory is a facility of the National Science Foundation operated under cooperative agreement by Associated Universities, Inc.

%\vspace{5mm}
\facilities{ALMA, VLT(SPHERE)}
 

\software{analysisUtils~(\url{https://casaguides.nrao.edu/index.php/Analysis_Utilities}), AstroPy~\citep{Astropy2018}, CASA~\citep{McMullin2007}, frank~\citep{Jennings2020}, matplotlib~\citep{matplotlib2007}, Scipy~\citep{scipy2020NatMe}}

\appendix
%\restartappendixnumbering
\twocolumngrid

\section{Frank Fitting Results} \label{sec:frank}
\begin{figure*}[!th]
\centering
    \includegraphics[width=0.9\textwidth, trim=50 0 100 50, clip]{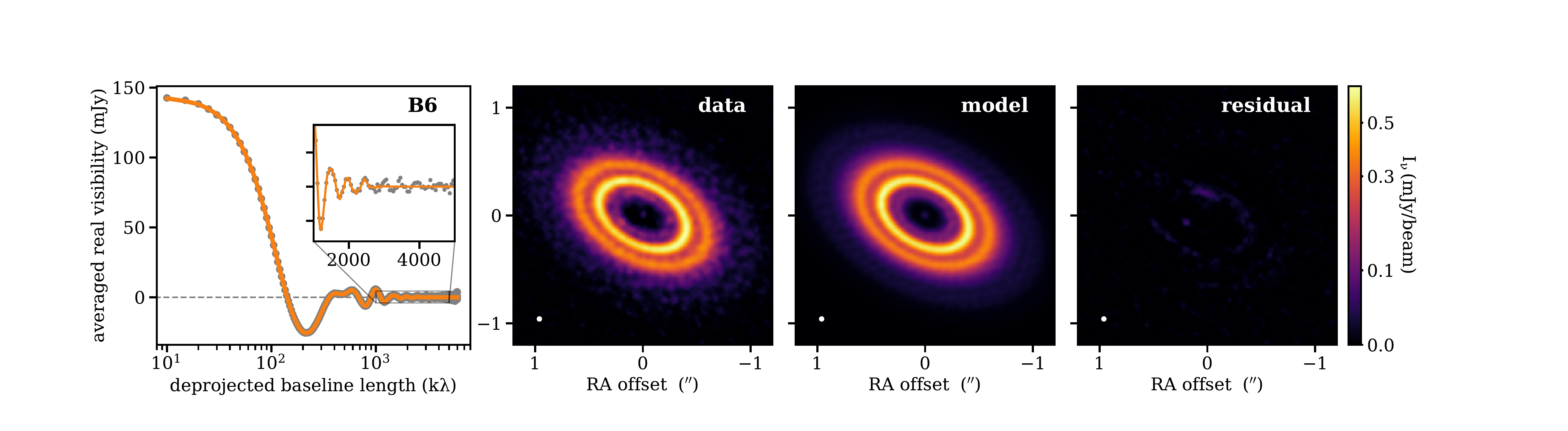} \\
    \includegraphics[width=0.9\textwidth, trim=50 0 100 50, clip]{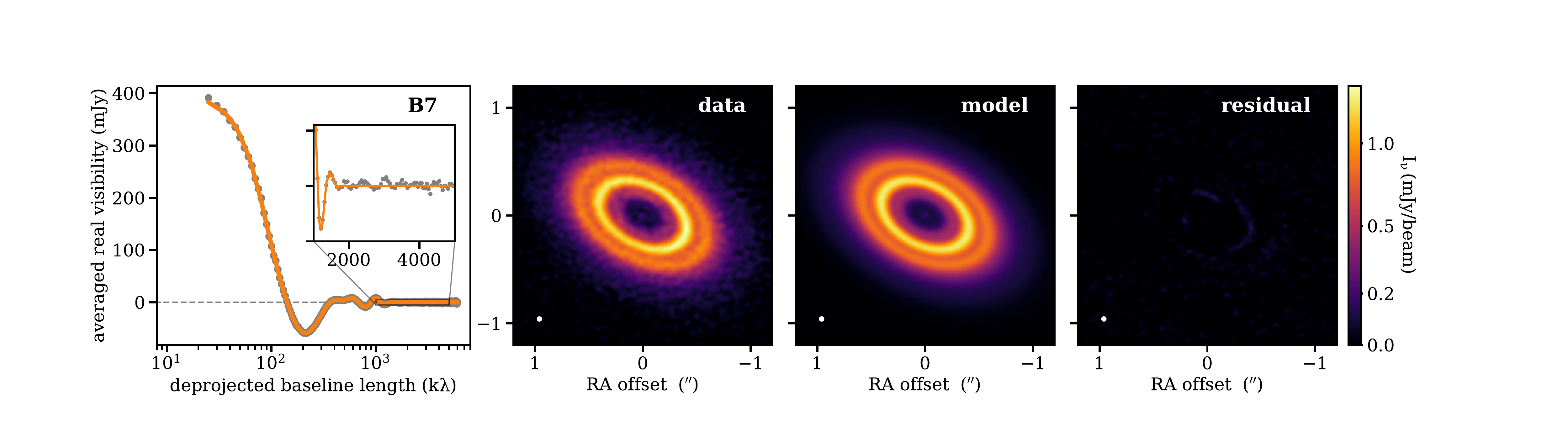} \\
\caption{Comparisons of the data and frank model in the visibility profile and image plane for B6 (upper panel) and B7 (lower panel). The data visibilities are binned in 5\,$k\lambda$ shown in grey points. The insert panel shows an zoom-in view of the long-baseline part outward of 1\,M$\lambda$ with data binned in 50\,$k\lambda$.  \label{fig:frank-comp}}
\end{figure*}

Figure~\ref{fig:frank-comp} compares the LkCa~15 ALMA data at B6 and B7 with their corresponding models obtained from \texttt{frank} fitting, in both the visibility profile and image plane. The model and residual images are constructed with identical parameters in the \texttt{tclean} task as for the data. As seen from the residual map, mm emission in the LkCa~15 disk can be broadly described by an axisymmetric model. 

\begin{figure}[!h]
\centering
    \includegraphics[width=0.45\textwidth]{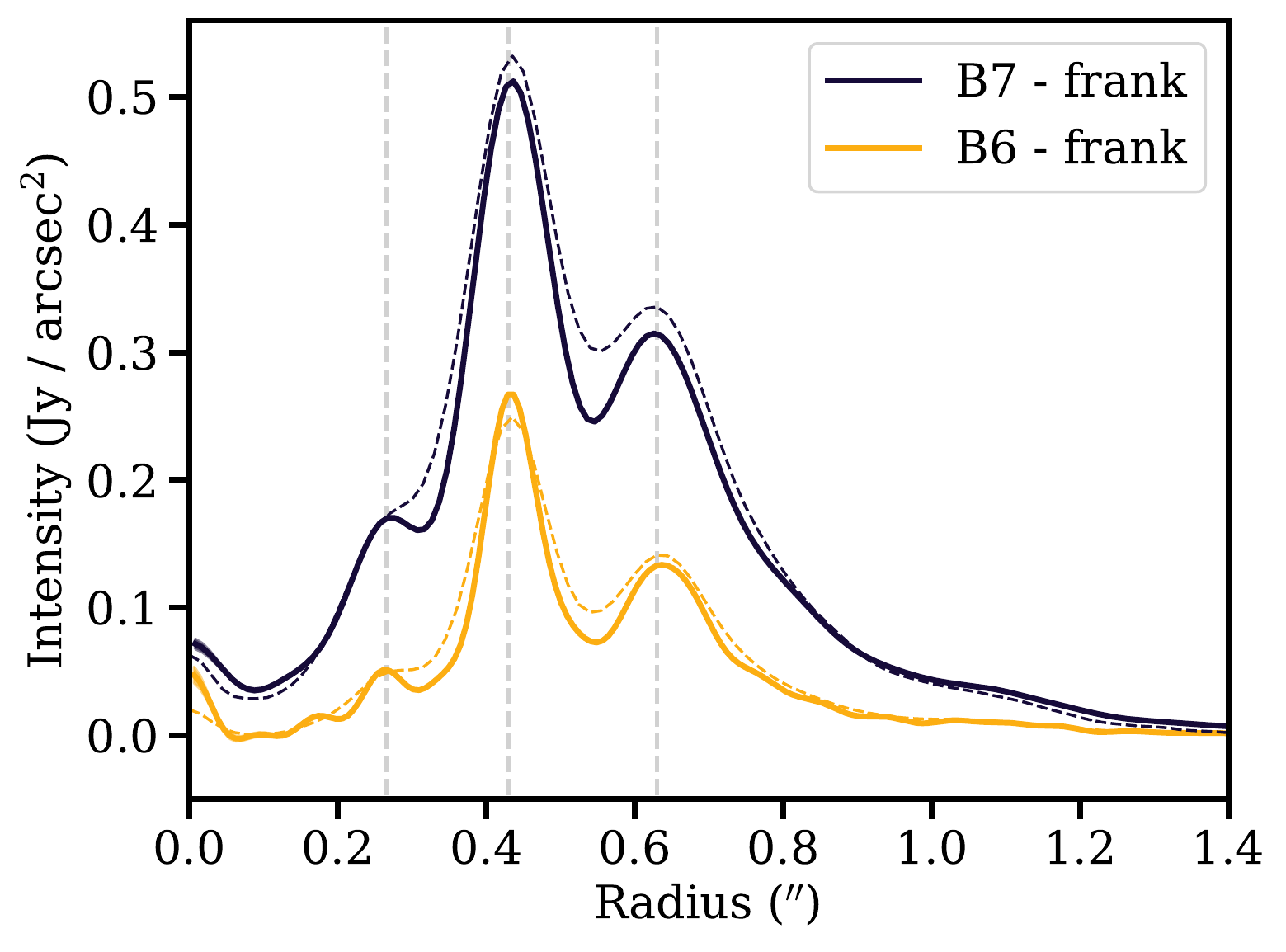} \\
    \caption{The model brightness profiles derived with \texttt{frank}, using the parameters described in Section~\ref{sec:results}. The three prominent dust rings are marked with the grey vertical lines. The data radial profiles from images convolved with a 50\,mas beam are shown in dashed lines for comparison.  \label{fig:frank-rp}}
\end{figure}

The model brightness profiles are shown in   Figure~\ref{fig:frank-rp}, and properties of the three prominent dust rings are summarized in Table~\ref{tab:frank}, each described by a Gaussian profile. Rings at the longer wavelength of 1.3\,mm are generally narrower. The notable difference of the width in the B42 ring is mainly subject to the retrieval of additional emission component in B6 (around $0\farcs18$). Beyond the three dust rings, the outer disk is associated with a broad emission halo, extending to at least $1\farcs2$ as seen from the B7 profile.

%%%%%%%%%%%%%%%%%%%%%%%%%%%%%%%%%
%%%%%%%%%%%%%%%%%%%%%%%%%%%%%%%%%
%%%%%%%%%%%%%%%%%%%%%%%%%%%%%%%%%
%%%%%%%%%%%%%%%%%%%%%%%%%%%%%%%%%

%%%%%%%%%%%%%%%%%%%%%%%%%%%%%%%%%
%%%%%%%%%%%%%%%%%%%%%%%%%%%%%%%%%
%%%%%%%%%%%%%%%%%%%%%%%%%%%%%%%%%
%%%%%%%%%%%%%%%%%%%%%%%%%%%%%%%%%

\begin{deluxetable*}{cccc|ccc|ccc}
%[!t]
%\rotate
\tabletypesize{\scriptsize}
\tablecaption{Dust Ring Properties from frank Fitting \label{tab:frank}}
\tablewidth{0pt}
\tablehead{
\colhead{Band} & \colhead{$A_{\rm B42}$} & \colhead{$r_{\rm B42}$} & \colhead{$\sigma_{\rm B42}$} & \colhead{$A_{\rm B69}$} & \colhead{$r_{\rm B69}$} & \colhead{$\sigma_{\rm B69}$} & \colhead{$A_{\rm B101}$} & \colhead{$r_{\rm B101}$} & \colhead{$\sigma_{\rm B101}$} \\ 
\cmidrule(lr){2-4} \cmidrule(lr){5-7} \cmidrule(lr){8-10} 
\colhead{} & \colhead{(log Jy Sr$^{-1}$)} & \colhead{(arcsec)} & \colhead{(arcsec)} & \colhead{(log Jy Sr$^{-1}$)} & \colhead{(arcsec)} & \colhead{(arcsec)} & \colhead{(log Jy Sr$^{-1}$)} & \colhead{(arcsec)} & \colhead{(arcsec)}  \\ 
}
\colnumbers
\startdata
B6 & $9.34\pm0.01$ & $0.266\pm0.001$ & $0.033\pm0.001$ & $10.04\pm0.01$ & $0.435\pm0.001$ & $0.043\pm0.001$ & $9.75\pm0.01$ & $0.638\pm0.001$ & $0.068\pm0.001$ \\
B7 & $9.81\pm0.02$ & $0.269\pm0.004$ & $0.064\pm0.007$ & $10.28\pm0.01$ & $0.430\pm0.001$ & $0.050\pm0.001$ & $10.09\pm0.01$ & $0.629\pm0.002$ & $0.107\pm0.003$ \\
\enddata
\tablecomments{Column 1: ALMA band. Column 2-4: the fitted amplitude, location, and Gaussian sigma for the B42 ring. Column 5-7: for the B69 ring. Column 8-10: for the B101 ring.}
\end{deluxetable*}

\section{Image Gallery} \label{sec:gallery}
\begin{figure*}[!th]
\centering
    \includegraphics[width=0.8\textwidth,]{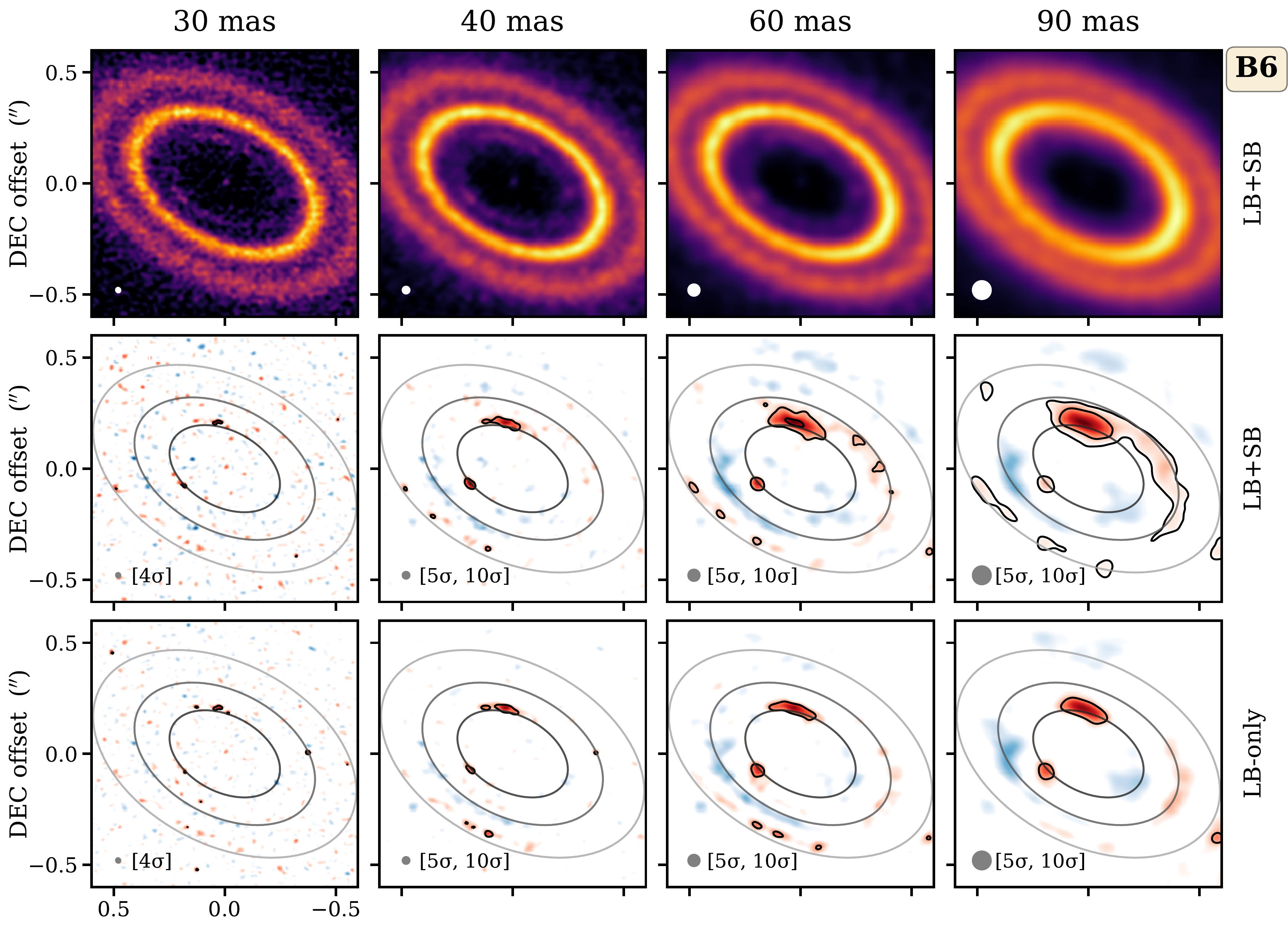} \\
    \includegraphics[width=0.8\textwidth,]{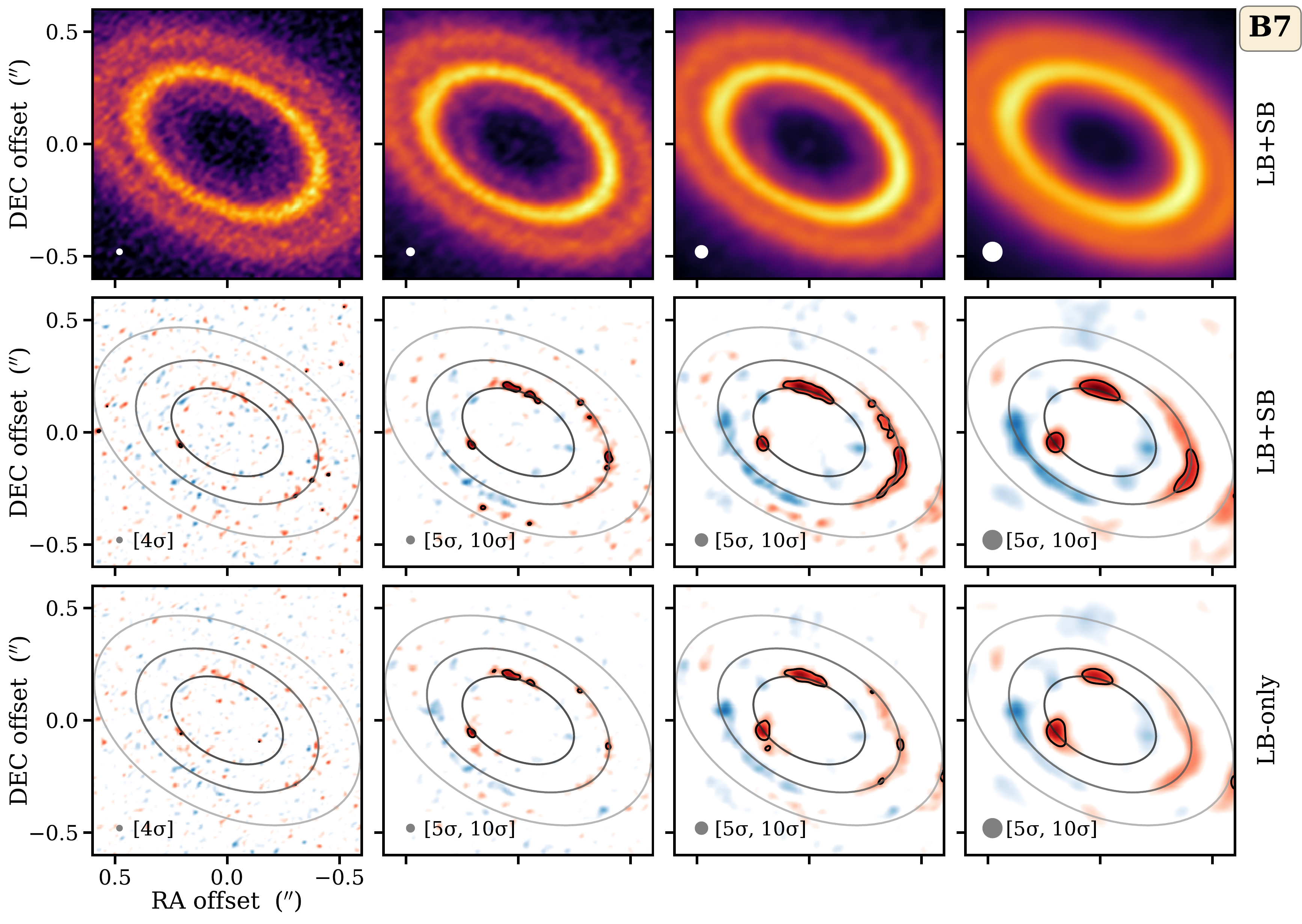} \\
\caption{The disk continuum images and \texttt{frank} residual maps with different synthesized beams for both B6 (upper panels) and B7 (lower panels). Columns correspond to different beam sizes, as indicated in the top panel. The disk images and first row of residual maps were created with the full combined data, while the second row used only the long-baseline data sets. 
 \label{fig:gallery}}
\end{figure*}

To assess how angular resolution may affect the recovery of the clump and arc features, we created a set of new images using different combinations of robust weighting parameters and uvtaper in \texttt{tclean}. The images, as well as their corresponding \texttt{frank} residual maps, are shown in Figure~\ref{fig:gallery}, spanning a range of beam sizes from 30 to 90\,mas. Except for the extreme case of 30\,mas where the B42 ring itself is only marginally detected, the two excess emission features are always clearly identified. However, with a large beam of 90\,mas, the B42 ring would be largely unresolved, thus the retrieval of feature properties (e.g., amplitude, shape) can be highly uncertain. 
The detection of the two features is also robust when only using the long-baseline configuration data, as shown in the bottom rows of Figure~\ref{fig:gallery}.

%%%%%%%%%%%%%%%%%%%%%%%%%%%%%%%%%
%%%%%%%%%%%%%%%%%%%%%%%%%%%%%%%%%
%%%%%%%%%%%%%%%%%%%%%%%%%%%%%%%%%
%%%%%%%%%%%%%%%%%%%%%%%%%%%%%%%%%

\bibliography{ms}{}
\bibliographystyle{aasjournal}

\end{CJK*}
\end{document}